\documentclass[12pt]{iopart}
\usepackage{iopams,graphicx}
\begin{document}

\title[Overlapping Unit Cells in 3-d Quasicrystal Structure]{Overlapping
Unit Cells in 3-d Quasicrystal Structure}

\author{Helen Au-Yang and Jacques H H Perk}

\address{Department of Physics, Oklahoma State University, 
145 Physical Sciences, Stillwater, OK 74078-3072, USA}
\ead{perk@okstate.edu}
\begin{abstract}
A 3-dimensional quasiperiodic lattice, with overlapping unit cells and
periodic in one direction, is constructed using grid and projection
methods pioneered by de Bruijn. Each unit cell consists of 26 points, of
which 22 are the vertices of a convex polytope $\cal P$, and 4 are
interior points also shared with other neighboring unit cells. Using
Kronecker's theorem the frequencies of all possible types of overlapping
are found.
\end{abstract}

\vspace{2pc}

\font\mymsbm=msbm10 at 12pt
\def\myb{\boldsymbol}
\def\br{{\myb R}}
\def\bd{{\myb d}}
\def\bk{{\myb k}}
\def\bD{{\myb D}}
\def\bw{{\myb w}}
\def\bW{{\myb W}}
\def\bP{{\myb P}}
\def\bQ{{\myb Q}}
\def\bn{{\myb n}}
\def\be{{\myb e}}
\def\T{{\rm T}}
\section{Introduction}
Since the startling discovery of five-fold symmetry in quasiperiodic
materials in 1984 \cite{SBGC}, a great deal of research has been done on
this subject by both physicists and mathematicians. Originally,
quasicrystals were constructed by filling the space aperiodically with
nonoverlapping tiles, such as in Penrose tilings
\cite{Penrose,GrSh,Bruijn1}\footnote{For a recent review of the
theory of quasicrystals, see, e.g., \cite{KrPa}.}. However, recently,
Gummelt
\cite{Gummelt} motivated by physical considerations, proposed a
description of quasicrystals in terms of overlappings of decagons. Further
research \cite{SJ,SJSTAT,LR,LRK,Kramer1,DuGr,GGA,ALS} has shown that this
may be a more sensible way to understand quasicrystalline materials---made
of overlapping unit cells sharing atoms of nearby neighbors \cite{SJSTAT}.
Here we use a multigrid method to produce a new example of 3-dimensional
overlapping unit cells.

\section{Multigrid construction}
It is well-known that a Penrose tiling can be obtained by the projection
of a slab of the 5-d euclidean lattice onto a particular 2-d plane 
${\cal D}$ \cite{Bruijn1,Bruijn2,GRh}, and hence its diffraction pattern
\cite{Elser,DK,Mackay2} has ten-fold symmetry. It is also known
that not all lattice points $\bk$ in ${\mbox{\mymsbm Z}}^5$ are allowed
(in the sense that they can be mapped onto vertices of a Penrose tiling);
only those points whose projections into the 3-dimensional orthogonal
space ${\cal W}$ are inside the window of acceptance \cite{Bruijn2,BJKS}
contribute. The window has been shown \cite {Bruijn2} to be the projection
of the 5-d unit cell Cu(5) with $2^5$ vertices into this 3-d space
$\cal W$. Each facet shared by two neighboring 5-d unit cell cubes is
4-dimensional and when projected into 3-d space it produces a polyhedron
$\cal K$ with 12 faces. Therefore, the projections of two adjacent 5-d
unit cells into 3d must share a common projected facet $\cal K$. Thus the
idea of overlapping decagons must have its extension to three dimension,
by projecting the 5-d lattice into the space $\cal W$.

If $\bd_j$ are the generators of the plane ${\cal D}$ and $\bw_j$ are the
generators of its orthogonal space ${\cal W}$, then the projection
operators are the matrices $\bD^\T=(\bd_0,\ldots,\bd_4)$ and
$\bW^\T=(\bw_0,\ldots,\bw_4)$, such that $\bD^\T\bW=\bW^\T\bD=0$, where
the superscript T denotes matrix transpose. More specifically, we choose
\begin{equation}
\bd^\T_j=(\cos j\theta,\sin j\theta),\quad
\bw^\T_j=(\cos 2j\theta,\sin 2j\theta,1)=(\bd^\T_{2j},1),
\label{generators}
\end{equation}
where $j=0,\ldots,4$ and $\theta=2\pi/5$. Using notations and ideas
introduced by de Bruijn \cite{Bruijn1}, we consider the five grids
consisting of bundles of equidistant planes defined by
\begin{equation}
x\cos 2j\theta+y\sin 2j\theta+z+\gamma_j=\bw^\T_j\br+\gamma_j=k_j,
\label{grid}
\end{equation}
for $\quad j=0,\ldots,4$, $k_j\in{\mbox{\mymsbm Z}}$. In (\ref{grid}),
$\br^\T=(x,y,z)$, and the $\gamma_j$ are real numbers which shift the
grids from the origin. We denote their sum by
\begin{equation}
\gamma_{0}+\gamma_{1}+\gamma_{2}+\gamma_{3}+\gamma_{4}=c.
\label{shiftc}
\end{equation}
Without loss of generality, we may restrict $c$ to $0\le c<1$.

It has been shown by de Bruijn \cite{Bruijn1} that the Penrose tiling
associated with a 2-d pentagrid has simple matching rules only for
$c=0$. For $0<c<1$ the corresponding generalized Penrose tilings do not
satisfy simple matching rules, and have seven different sets of vertices
corresponding to the different intervals of $c$ \cite{APwindow,APquasi}.
Nevertheless, the diffraction patterns are believed to be the same
for all values of $c$ \cite {LSt,SoSt,IsYa}.
 
Let the integer $k_j$ be assigned to all points sandwiched between the
grid planes defined by $k_j-1$ and $k_j$. Then, five integers 
\begin{equation}
K_j(\br)=\lceil\bw^\T_j\br+\gamma_j\rceil,\quad j=0,\ldots,4,
\label{mesh}
\end{equation}
with $\lceil x\rceil$ the smallest integer greater than or equal to $x$,
are uniquely assigned to every point $\br$ in ${\mbox{\mymsbm R}}^3$.
A mesh in ${\mbox{\mymsbm R}}^3$ is now an interior volume, enclosed by
grid planes, containing points with the same five integers. One next
maps each mesh to a vertex in $\cal W$ by 
\begin{equation}
\fl
{\myb g}(\br)=\sum_{j=0}^4 K_j(\br)\bw_j=\bW^\T{\myb K}(\br),\quad
{\myb K}^\T(\br)=(K_0(\br),\ldots,K_4(\br)).
\label{map}\end{equation}
The resulting collection of vertices
${\cal L}=\{{\myb g}(\br)|\br\in{\mbox{\mymsbm R}}^3\}$ is a
3-dimensional aperiodic lattice. It has been proven by de Bruijn
\cite{Bruijn2}, that a point $\bk$ in ${\mbox{\mymsbm Z}}^5$ satisfies
the so-called {\em mesh condition} and therefore can be mapped into
${\cal L}$ if and only if $\bD^\T(\bk-{\myb\gamma})=\bD^\T{\myb\lambda}$,
where ${\myb\gamma}^\T=(\gamma_0,\ldots,\gamma_4)$ and 
${\myb\lambda}^\T=(\lambda_0,\ldots,\lambda_4)$ with $0<\lambda_j<1$ so
that ${\myb\lambda}$ is a point inside the 5-d unit cube Cu(5). Thus, the
{\em window of acceptance} is the interior of the convex hull of the
points $\bD^\T{\bn_i}$, where $\bn_i$ are the $2^5$ vertices of the 5-d
unit cube Cu(5), see figure~1. We choose the 32 $\bn_i$'s as follows
\begin{eqnarray}
&&\bn_0=0,\quad\bn_{31}=\be_0+\be_1+\be_2+\be_3+\be_4,\nonumber\\
&&\bn_{j+1}=\be_{1-2j},\quad
\bn_{j+6}=\be_{1-2j}+\be_{4-2j},\quad
\bn_{j+11}=\be_{1-2j}+\be_{2-2j},\nonumber\\
&&\bn_{j+16}=\be_{1-2j}+\be_{2-2j}+\be_{5-2j},\quad
\bn_{j+21}=\be_{1-2j}+\be_{2-2j}+\be_{4-2j},\nonumber\\
&&\bn_{j+26}=\be_{1-2j}+\be_{2-2j}+\be_{4-2j}+\be_{5-2j},\quad
(j=0,\ldots,4),
\label{howto}
\end{eqnarray}
where $\be_0,\ldots,\be_4$ are the standard unit vectors in
${\mbox{\mymsbm R}}^5$, with subscripts counted mod 5
($\be_j\equiv\be_{j\pm5}$).
\begin{figure}[tbh]
\begin{center}
\includegraphics[width=0.7\hsize]{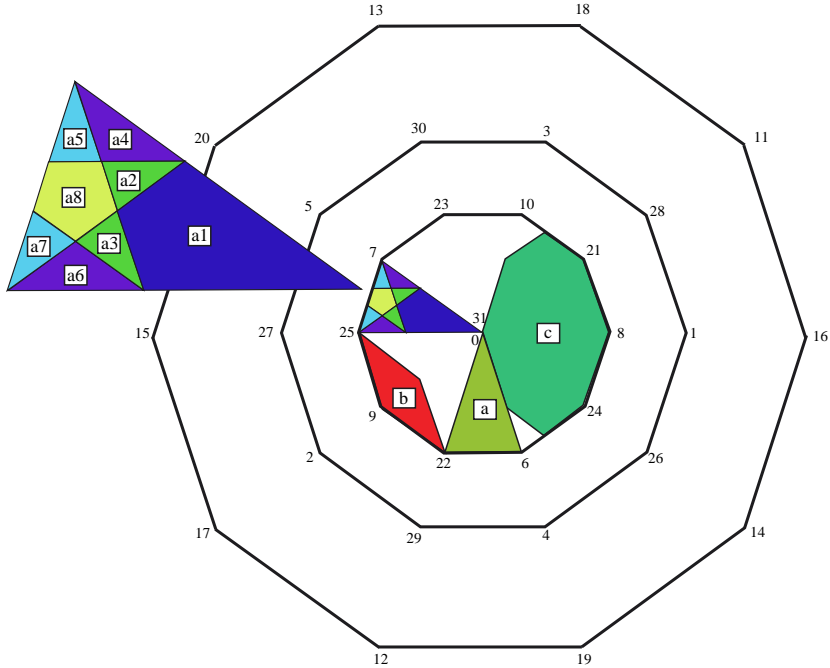}
\end{center}

\caption{The projection of the 5-d unit cube Cu(5) into the
orthogonal 2-d space $\cal D$. The window of acceptance is the interior of
outer decagon $\cal Q$ whose vertices are given by (\ref{decagonout}).
The innermost decagon is denoted by $\hat{\cal Q}$ and the middle decagon
by $\bar{\cal Q}$ with vertices given by (\ref{decagonin}). There are
ten triangles of type (a), which are all further subdivided in eight
regions of type (a1$),\ldots,($a8), as is indicated for one case in the
magnification on the left. This is determined by possible overlaps of
this triangle with the ten regions each of types (b) and (c), see text.}
\end{figure}

The projection of Cu(5) with these 32 points into $\cal W$ is a polytope
$\cal P$ with 40 edges connecting the 22 vertices, and with 20 faces. We
let $\bP_i=\bW^\T\bn_i$ for $i=0,\ldots,31$. The bottom is $\bP_0=(0,0,0)$
and top is $\bP_{31}=(0,0,5)$; they are called the tips of the polytope.
The remaining twenty vertices of $\cal P$ are 
\begin{eqnarray}
&\bP_{j+1}=(\bd_j,1),\quad \bP_{j+6}=(\bd_j+\bd_{j+1},2),\nonumber\\
&\bP_{j+21}=(-\bd_{j-2}-\bd_{j-1},3),\quad \bP_{j+26}=(-\bd_{j-1},4),
\label{hpoly}
\end{eqnarray}
for $j=0,\ldots,4$. The other 10 points $\bP_{11},\ldots,\bP_{20}$
are in the interior of the polytope and are given by
\begin{equation}
\bP_{11+j}=(\bd_j+\bd_{j+2},2),\quad \bP_{16+j}=(-\bd_{j+1}-\bd_{j-1},3),
\label{decagon}\end{equation}
again for $j=0,\ldots,4$.

The orthogonal projection of Cu(5) into $\cal D$ is a decagon $\cal Q$
with 10 edges connecting the 10 vertices, see figure~1. Let
$\bQ_i=\bD^\T\bn_i$ for $i=0,\ldots,31$. Then the vertices of the decagon
are
\begin{equation}
\bQ_{11+j}=-p\,\bd_{3-2j},\quad \bQ_{16+j}=p\,\bd_{5-2j},
\label{decagonout}
\end{equation}
with $j=0,\ldots,4$, and $p=(\sqrt5+1)/2$. The remaining 22 points
$\bQ_{0},\ldots,\bQ_{10}$ and $\bQ_{21},\ldots,\bQ_{31}$ are in the
interior; they are given by
\begin{eqnarray}
&\bQ_{0}=\bQ_{31}=0,\quad \bQ_{j+1}=\bd_{5-2j},\quad
\bQ_{26+j}=-\bd_{2-2j},\nonumber\\
&\bQ_{j+6}=p^{-1}\bd_{4-2j},\quad \bQ_{21+j}=-p^{-1}\bd_{3-2j}.
\label{decagonin}
\end{eqnarray}
Thus if the orthogonal projection $D^\T(\bk-{\myb\gamma})$ is in $\cal Q$,
then its projection $\bW^\T\bk$ is in ${\cal L}$.

Consider a polytope $\cal P$ in ${\cal L}$ whose bottom is the
projection of a point $\hat\bk$ which is orthogonally projected onto the
center of the decagon, then the ten vertices of $\cal P$ at height
$z=1$ or $z=4$ correspond to the vertices of the middle decagon
$\bar{\cal Q}$, see figure~1; the ten vertices of the polytope at height
$z=2$ or $z=3$, correspond to the vertices of the innermost decagon
$\hat{\cal Q}$; while the ten interior points of the polytope correspond
to the vertices of the (outer) decagon ${\cal Q}$ shown in figure~1.

The points in the quasiperiodic lattice $\cal L$ need careful analysis. We
shall show that $\cal L$ consists of polytopes $\cal P$ with 22 vertices
and 4 interior points which are shared with neighbouring polytopes. We
shall also show that it is periodic in the $z$-direction---the direction
of the line joining the two tips of $\cal P$---with period 5 corresponding
to the height of $\cal P$.

\section{Overlapping unit cells}
The window $\cal Q$ is known \cite{Bruijn1,Bruijn3} to be everywhere dense
and uniformly distributed. Thus each point in $\cal Q$ corresponds to
a point in the quasiperiodic lattice $\cal L$. Using an idea of de Bruijn
\cite{Bruijn1}, we may find out the condition for both $\bk$ and
$\bk+\bn_i$ to satisfy the mesh condition---to lie both in the window of
acceptance, which give insights to how points in $\cal L$ are related to
one another.

It is straightforward to show that every point inside the innermost
decagon $\hat{\cal Q}$ corresponds to a point in $\cal L$ that is
connected with 10 neighbors, and is in fact a tip of a polytope, as the
middle decagon $\bar{\cal Q}$ whose center is shifted to a point inside
$\hat{\cal Q}$ still lies inside $\cal Q$. This innermost decagon
$\hat{\cal Q}$ is further divided into 10 triangles \footnote{A rotation
of points in $\hat{\cal Q}$ by an angle of $2\pi\ell/5$, corresponds to a
rotation of the polytopes in $\cal L$ by an angle of $4\pi\ell/5$ about
the $z$-axis, while a rotation by $\pi$---inversion of points in
$\hat{\cal Q}$ through the origin of decagon $\hat{\cal Q}$---corresponds
to the 3-d inversion of polytopes in $\cal L$ through the center of each
polytope.}. Whenever the center of a decagon $\cal Q$ is shifted to a
point inside one of the triangles, four of the vertices of the shifted
outer decagon now lie inside $\cal Q$. More precisely, if
$\bD^\T(\bk-{\myb \gamma})$ is inside triangle (a) in figure~1, then
$\bW^\T\bk$ is a tip of a polytope $\cal P$ whose interior points
$\bW^\T(\bk+\bn_{20})$, $\bW^\T(\bk+\bn_{13})$, $\bW^\T(\bk+\bn_{18})$ and
$\bW^\T(\bk+\bn_{11})$---corresponding to the four points on $\cal Q$ on
the opposite side of triangle (a)---are now also in $\cal L$. This means
each polytope in ${\cal L}$ can have only four interior points which are
also in ${\cal L}$. Thus each such unit cell contains 26 atoms, 22
exterior, and 4 interior sites. 

It is also easy to find out how the polytopes share these interior points.
This is equivalent to finding the condition that both $\bk$ and
$\bk+\bn_{i}$ for $i=1,\ldots,10$ or $i=21,\ldots,30$ are
orthogonally projected into $\hat{\cal Q}$.

Consider the ten vertices of the polytope at height $z=1$ or $z=4$
corresponding to the vertices of middle decagon $\bar{\cal Q}$. When the
center of the decagon is shifted to a point in one of ten rhombs of type
(b) shown in figure~1, then one of the ten vertices of the shifted
$\bar{\cal Q}$ is inside $\hat{\cal Q}$. For the example in figure~1,
$\bW^\T(\bk)$ and $\bW^\T(\bk+\bn_{28})$ are both tips of polytopes, with
$\bn_{28}$ the point on $\bar{\cal Q}$ on the opposite side of rhomb (b).
As a consequence, the polytope whose top is $\bW^\T(\bk+\bn_{28})$ shares
with the polytope $\cal P$ whose bottom is $\bW^\T(\bk)$ a polyhedron
$\cal K$ with 12 faces.

Moreover, the vertices of the polytope at $z=2$ or $z=3$ correspond to
the vertices of $\hat{\cal Q}$. When the center of this decagon is shifted
to a point in one of ten octagons of type (c) shown also in figure~1, then
one of the ten vertices of the shifted $\hat{\cal Q}$ moves inside
$\hat{\cal Q}$. For the example in figure~1, $\bW^\T(\bk)$ and
$\bW^\T(\bk+\bn_{25})$ are then both tips of polytopes, while $\bn_{25}$
is the point on $\hat{\cal Q}$ on the side opposite to the octagon. Such
two polytopes share a polyhedron $\cal J$ with 6 faces. When vertices of
$\cal P$ at both $z=3$ and $z=4$ are inside $\hat{\cal Q}$, they are tops
of some polytopes\footnote{We shall later show that lattice ${\cal L}$
is periodic in the $z$-direction with period 5.} that share with
${\cal P}$ either a polyhedron ${\cal K}$ or ${\cal J}$.

By considering how these twenty regions intersect, we find that each of
the triangles in $\hat{\cal Q}$ is further divided into eight regions,
as shown for one case in figure~1, with this triangle magnified on the
left. When the orthogonal projection of $\bk-{\myb \gamma}$ is in (a1),
its projection into $\cal W$ is a polytope $\cal P$ intersecting with four
others whose tips are at $\bW^\T(\bk+\bn_{21})$, $\bW^\T(\bk+\bn_{8})$,
$\bW^\T(\bk+\bn_{24})$ and $\bW^\T(\bk+\bn_{6})$, and sharing with each
of them a polyhedron of type $\cal J$. When the projection is in (a2) or
(a3), either $\bW^\T(\bk+\bn_{26})$ or $\bW^\T(\bk+\bn_{1})$ becomes also
a tip of a polytope, so that $\cal P$ intersects with five polytopes,
sharing with one of them a polyhedron of type $\cal K$. In (a8),
$\cal P$ intersects with all of the above six polytopes. In regions
(a5) or (a7), $\bW^\T(\bk+\bn_{21})$ or $\bW^\T(\bk+\bn_{6})$ is no
longer a tip and $\cal P$ intersects with five polytopes sharing with two
of them a polyhedron of type $\cal K$. Finally, while in (a4) or (a6),
either $\bW^\T(\bk+\bn_{1})$ or $\bW^\T(\bk+\bn_{26})$ is no longer a tip,
and $\cal P$ intersects with four polytopes sharing with one of them
a polyhedron $\cal K$ and with the other three a $\cal J$.

Therefore, in summary, there are only five different possibilities:
\begin{enumerate}
\item{The points inside a quadrilateral of type (a1) correspond to a
polytope intersecting with four other polytopes sharing with each a
polyhedron of type $\cal J$. An example of this case is shown in
figure~2a.}
\item{Points inside triangles of type (a2) or (a3) correspond to a
polytope intersecting with five other polytopes, sharing with one of them
a polyhedron $\cal K$ and with the other four polyhedra of type $\cal J$.
Such a case is shown in figure~2b.}
\item{If the point is inside triangles of type (a4) or (a6), the polytope
intersects with four other polytopes sharing with one of them a polyhedron
$\cal K$ and with the other three polyhedra of type $\cal J$.}
\item{If the point
is in a triangle (a5) or (a7), the polytope intersects with five other
polytopes sharing with two of them polyhedra of type $\cal K$ and with the
other three polyhedra $\cal J$.}
\item{Finally, if the point is inside a pentagon
(a8), the polytope intersects with six other polytopes sharing with two
of them a $\cal K$ and with the other four a $\cal J$, as is shown in
figure 3.}
\end{enumerate}
The relative frequencies are related to the ratios of their
areas and therefore the normalized probabilities are given by
\begin{eqnarray}
&P_{\rm a1}=2p^{-3},\quad P_{\rm a2}=P_{\rm a3}=p^{-6},\quad 
P_{\rm a4}=P_{\rm a6}=p^{-5},\nonumber\\
&P_{\rm a5}=P_{\rm a7}=p^{-6},\quad P_{\rm a8}=p^{-5}+p^{-7}.
\label{prob}
\end{eqnarray}
\begin{figure}[tbh]
\begin{center}
\includegraphics[width=0.4\hsize]{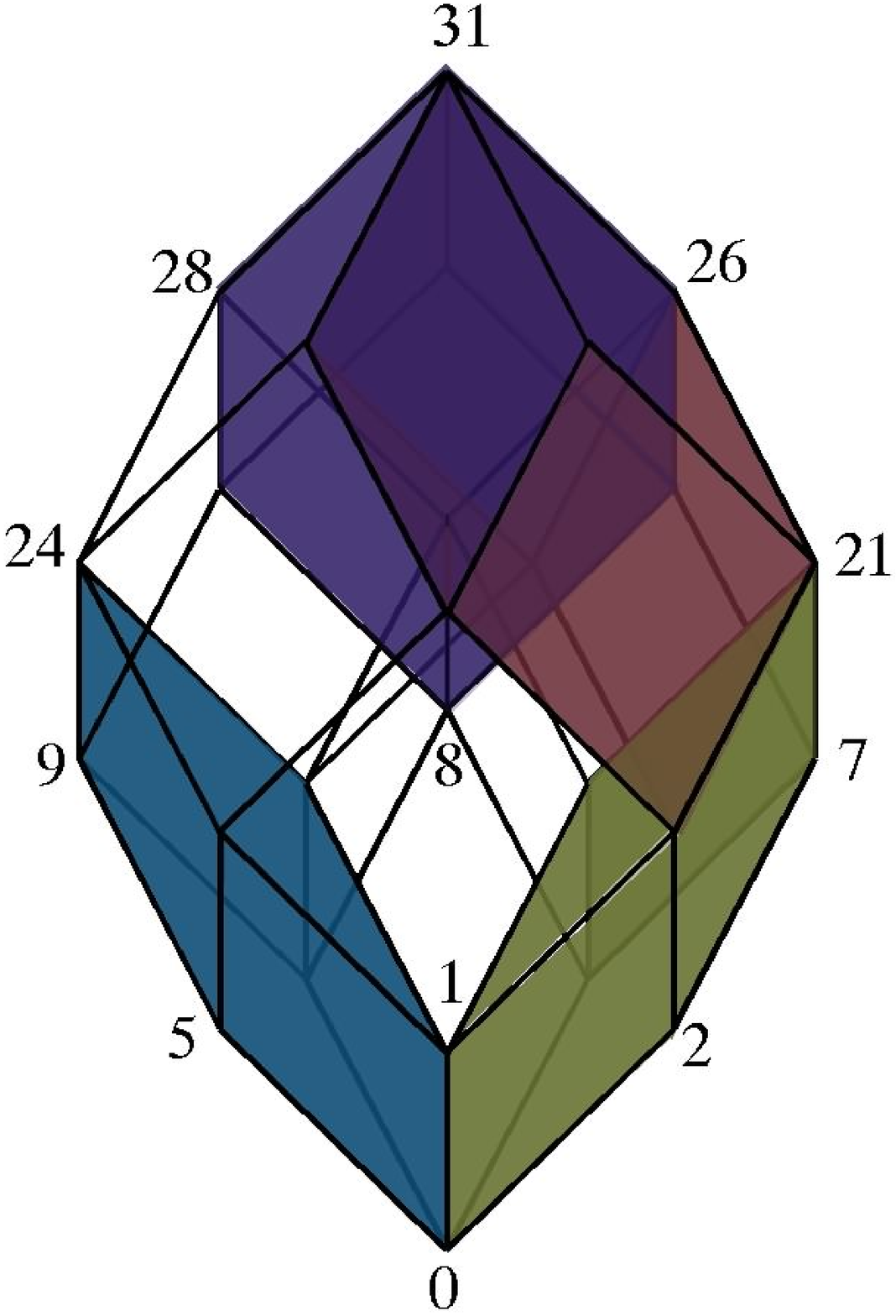}\hfil
\includegraphics[width=0.4\hsize]{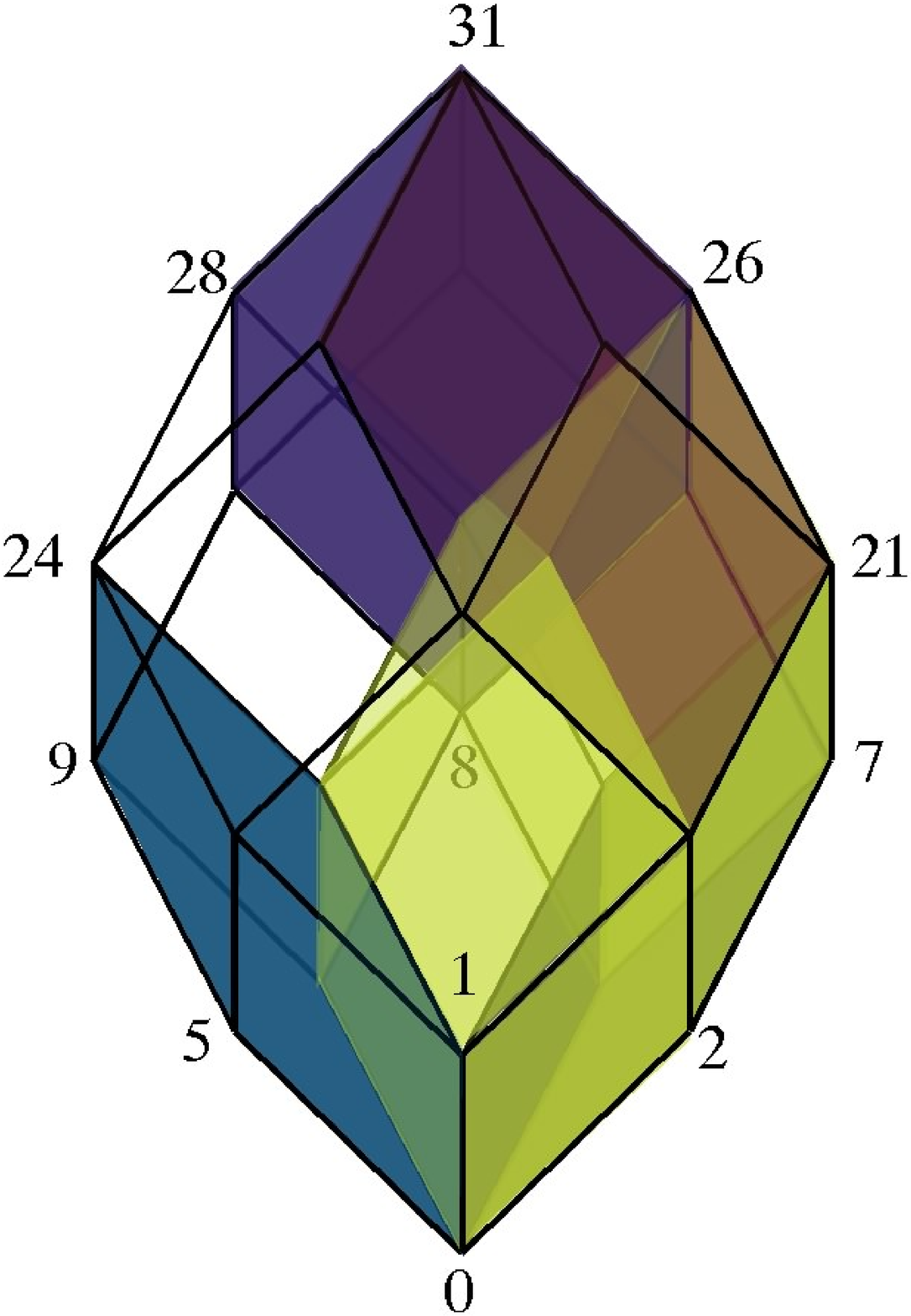}
\end{center}
\vskip6pt
\hbox to\hsize{\hspace*{12pt}\footnotesize
\hfil (a) \hfil\hfil (b)\hfil\hspace*{-2pt}}

\caption{(a) When the orthogonal projection of $\bk-{\myb \gamma}$ is in
region (a1), its projection in $\cal L$ is a polytope sharing an interior
point with each of four neighboring polytopes whose tips are projections
of $\bk+\bn_{21}$, $\bk+\bn_{8}$, $\bk+\bn_{24}$ and $\bk+\bn_{6}$; the
shared interior points are projections of $\bk+\bn_{11}$, $\bk+\bn_{16}$,
$\bk+\bn_{14}$ and $\bk+\bn_{19}$. (b) When it is in region (a2), the
polytope has five neighboring polytopes sharing the same interior points.
The additional polytope has its top at $\bW^\T(\bk+\bn_{26})$.}
\end{figure}
\begin{figure}[tbh]
\begin{center}
\includegraphics[width=0.4\hsize]{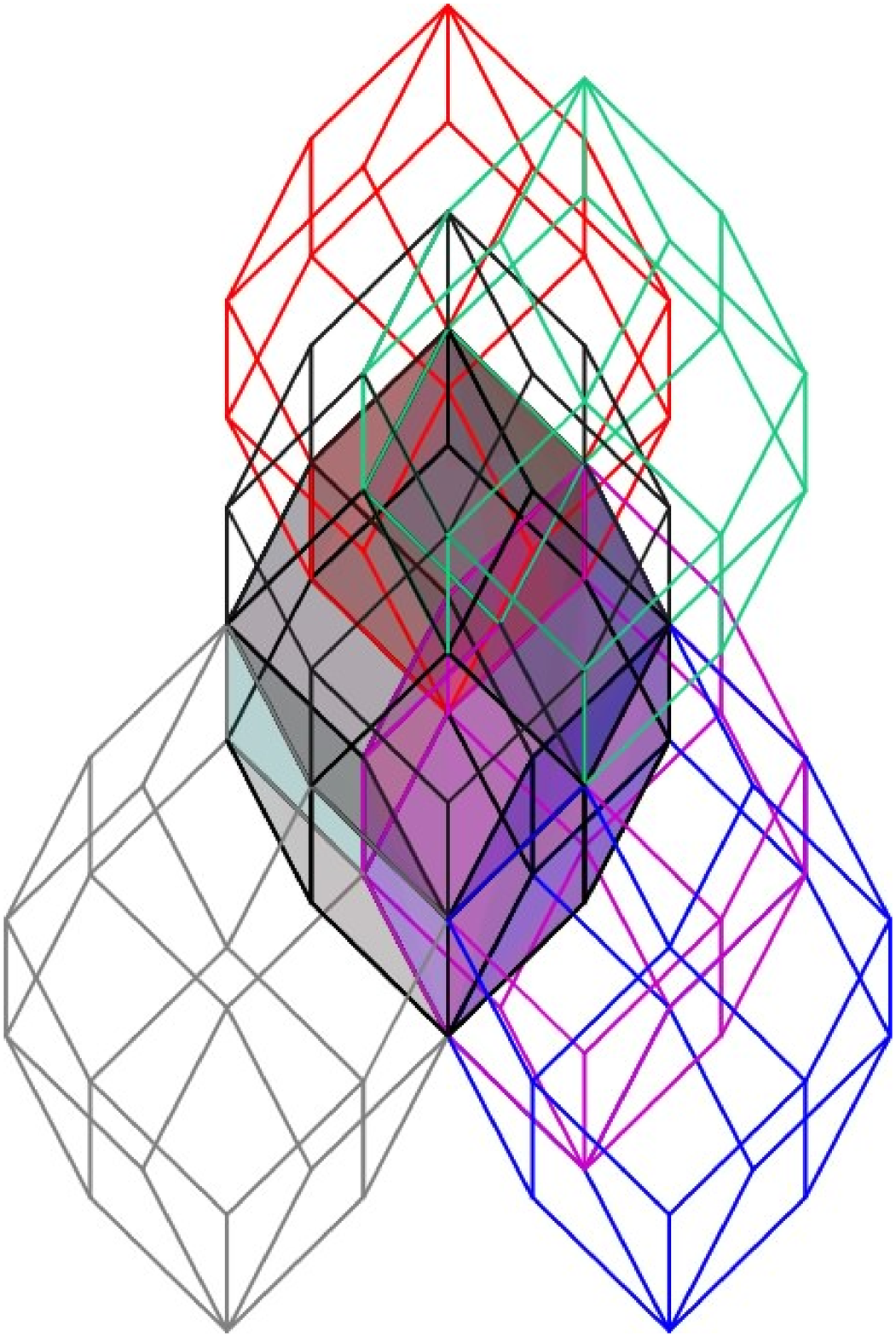}\hfil
\includegraphics[width=0.4\hsize]{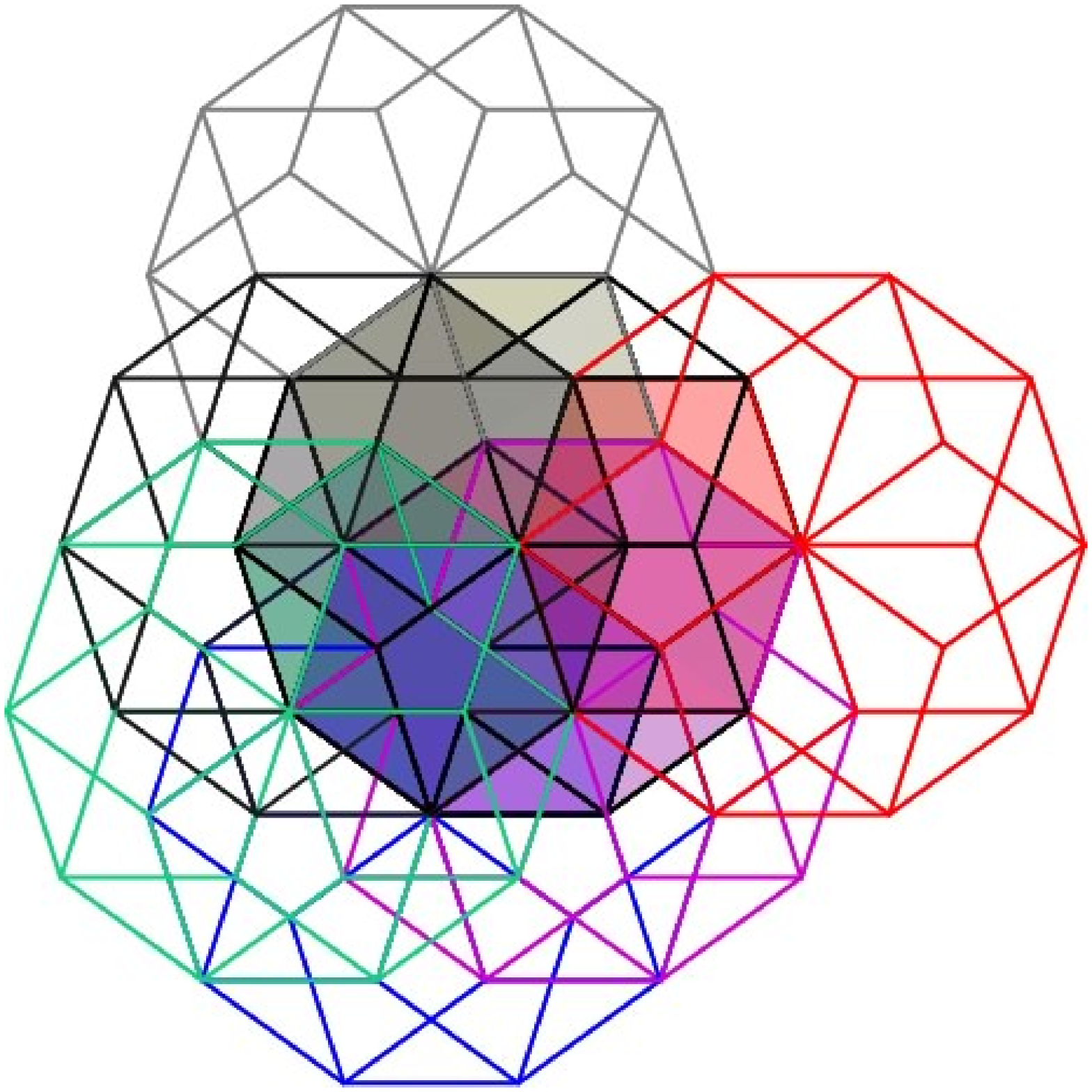}
\end{center}
\vskip6pt
\hbox to\hsize{\hspace*{12pt}\footnotesize
\hfil (a) \hfil\hfil (b)\hfil\hspace*{-2pt}}

\caption{The polytope $\cal P$ and its six neighbors in $\cal L$ for case
(a8). Their intersections with $\cal P$ are indicated with different
colors. In (a) the orientation is $(\theta,\phi)=(0,80^{\circ})$; in (b)
$(\theta,\phi)=(90^{\circ},0)$, so that each polytope appears as a
decagon.}
\end{figure}

\section{Parallelepipeds and windows}
Now the condition that $\bk$ satisfies the mesh condition can be further
simplified by considering parallelepiped $P(k_4,k_0,k_1)$ sandwiched
between the six grid planes $k_4-1$, $k_4$, $k_0-1$, $k_0$, $k_1-1$, 
and $k_1$. Using (\ref{grid}) for the grid planes, we can solve for the
points of intersections, and find that for every point $\br$ in
$P(k_4,k_0,k_1)$, we may write
\begin{eqnarray}
&K_0(\br)=k_0,\quad K_1(\br)=k_1,\quad
K_2(\br)=\lfloor\alpha\rfloor+k_4+m,\nonumber\\
&\, K_3(\br)=\lfloor\beta\rfloor+k_1+n,\quad K_4(\br)=k_4,
\label{piped}
\end{eqnarray}
where $\lfloor x\rfloor$ is the greatest integer less or equal to $x$,
while $m$ and $n$ are integers satisfying $-1\le m,n\le 2$, and
\begin{eqnarray}
&&\alpha=p^{-1}(k_0-k_1-\gamma_0+\gamma_1)+\gamma_2-\gamma_4,\nonumber\\
&&\beta=p^{-1}(k_0-k_4-\gamma_0+\gamma_4)+\gamma_3-\gamma_1.
\label{alpha}
\end{eqnarray}
The range of $m$ and $n$ in equation (\ref{piped}) is limited by their
possible values at the eight corners of the parallelepiped, but not all
16 choices are allowed by the mesh condition. Therefore, we now project
${\myb K}(\br)$ given by (\ref{piped}) to $\cal D$ and find
\begin{equation}
\bD^\T({\myb K}(\br)-{\myb \gamma})=\sum_{j=0}^4
(K_j(\br)-\gamma_j)\bd_j=(m-a)\bd_2+(n-b)\bd_3,
\label{window}
\end{equation}
in which $a\equiv\{\alpha\}\equiv\alpha-\lfloor\alpha\rfloor$ and
$b\equiv\{\beta\}\equiv\beta-\lfloor\beta\rfloor$. The vector
(\ref{window}) must lie within the decagon $\cal Q$. Thus the allowed
values of $m$ and $n$ are determined by $a$ and $b$ only. As the
differences $k_0-k_4$ and $k_0-k_1$ run through all integer values, we
find from Kronecker's theorem \cite{HW} that $a$ and $b$ are everywhere
dense and uniformly distributed in the interval $(0,1)$. Proofs of such
``ergodicity" in more general situations can be found, e.g., in the works
of Hof and Schlottmann \cite{Hof,Schlott}.

\section{Periodicity in third direction}
Furthermore, because of the difference property shown in (\ref{alpha}),
$\alpha$ and $\beta$, which determine the configuration of the
parallelepiped, remain the same, if
$(k_0,k_1,k_4)\to(k_0+\ell,k_1+\ell,k_4+\ell)$. As a consequence we find
${\myb K}(\br)\to{\myb K}(\br)+\ell\bn_{31}$, and its projection is
periodic in the $z$ direction with period equal to 5. It is also
interesting to note that if
${\myb \gamma}\to{\myb \gamma}-{\frac15}c\,\bn_{31}$, $\alpha$ and $\beta$
in (\ref{alpha}) are also unchanged, so that the projection into the
3d space does not show drastic changes when $c\ne0$, which is
behavior very different from the 2-d case \cite{APwindow}. 

It is not difficult to find values of $(a-m,b-n)$ in
(\ref{window}) corresponding to the ten vertices of the decagon $\cal Q$
given by (\ref{decagonout}). We find
\begin{eqnarray}
\begin{array}{ll}
\bQ_{16}\leftrightarrow (a-0,b-0)=(1,1),
  &\bQ_{11}\leftrightarrow (a-0,b+1)=(0,p),\cr
\bQ_{18}\leftrightarrow (a-1,b+1)=(-1,p),
  &\bQ_{13}\leftrightarrow (a-2,b+1)=(-p,1),\cr
\bQ_{20}\leftrightarrow (a-2,b-0)=(-p,0), 
 &\bQ_{15}\leftrightarrow (a-1,b-1)=(0,-1),\cr
\bQ_{17}\leftrightarrow (a-1,b-2)=(0,-p),
  &\bQ_{12}\leftrightarrow (a-0,b-2)=(1,-p),\cr
\bQ_{19}\leftrightarrow (a+1,b-2)=(p,-1),
  &\bQ_{14}\leftrightarrow (a+1,b-1)=(p,0),
\end{array}
\label{decagonval}
\end{eqnarray}
up to ambiguities when $a$ or $b$ is integer, as $m$ or $n$ changes by 1
when choosing $a$ or $b$ to be 0 or 1. The edges of decagon $\cal Q$ lead
to linear equations in $(a-m,b-n)$ and the mesh condition for
${\myb K}(\br)$ becomes a set of inequalities in $a-m$ and $b-n$ as shown
in figure 4. Hence, it is very easy to create a routine to generate
$\cal L$.
\begin{figure}[tbh]
\begin{center}
\includegraphics[width=0.5\hsize]{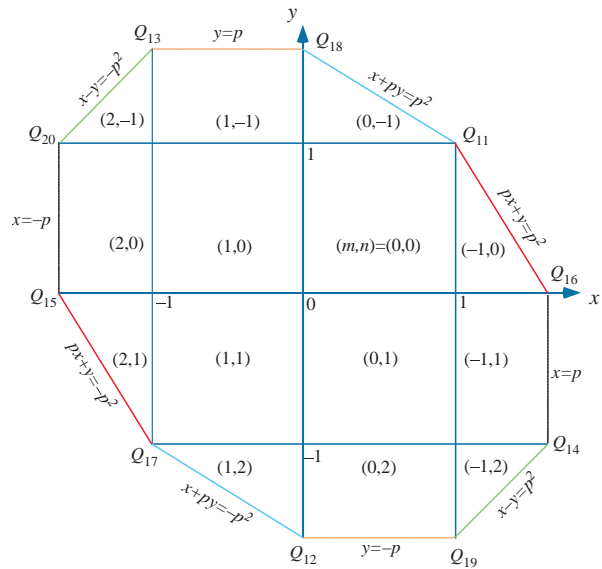}
\end{center}

\caption{The image of $\cal Q$ under the inverse mapping of (\ref{window})
into the $(a-m,b-n)$-plane. The boundaries of regions with different
$(m,n)$ are given by linear equations in terms of $x\equiv a-m$ and
$y\equiv b-n$.}
\end{figure}

We note that $\cal L$ is periodic in the longitudinal or $z$-direction
with period 5, and aperiodic in the horizontal directions. This is a model
with five layers which repeat in the $z$-direction periodically. In each
of the layers, the points behave similarly, and the allowed sites for
$z=3$ are shown in figure 5$\,$(a). This shows decagonal symmetry, but the
layer may not be covered by Penrose tiles. In figure 5$\,$(b), we have
sites for atoms in all five layers plotted, with sites in different layers
represented by different symbols and colors. 
\begin{figure}[tbh]
\begin{center}
\includegraphics[width=0.4\hsize]{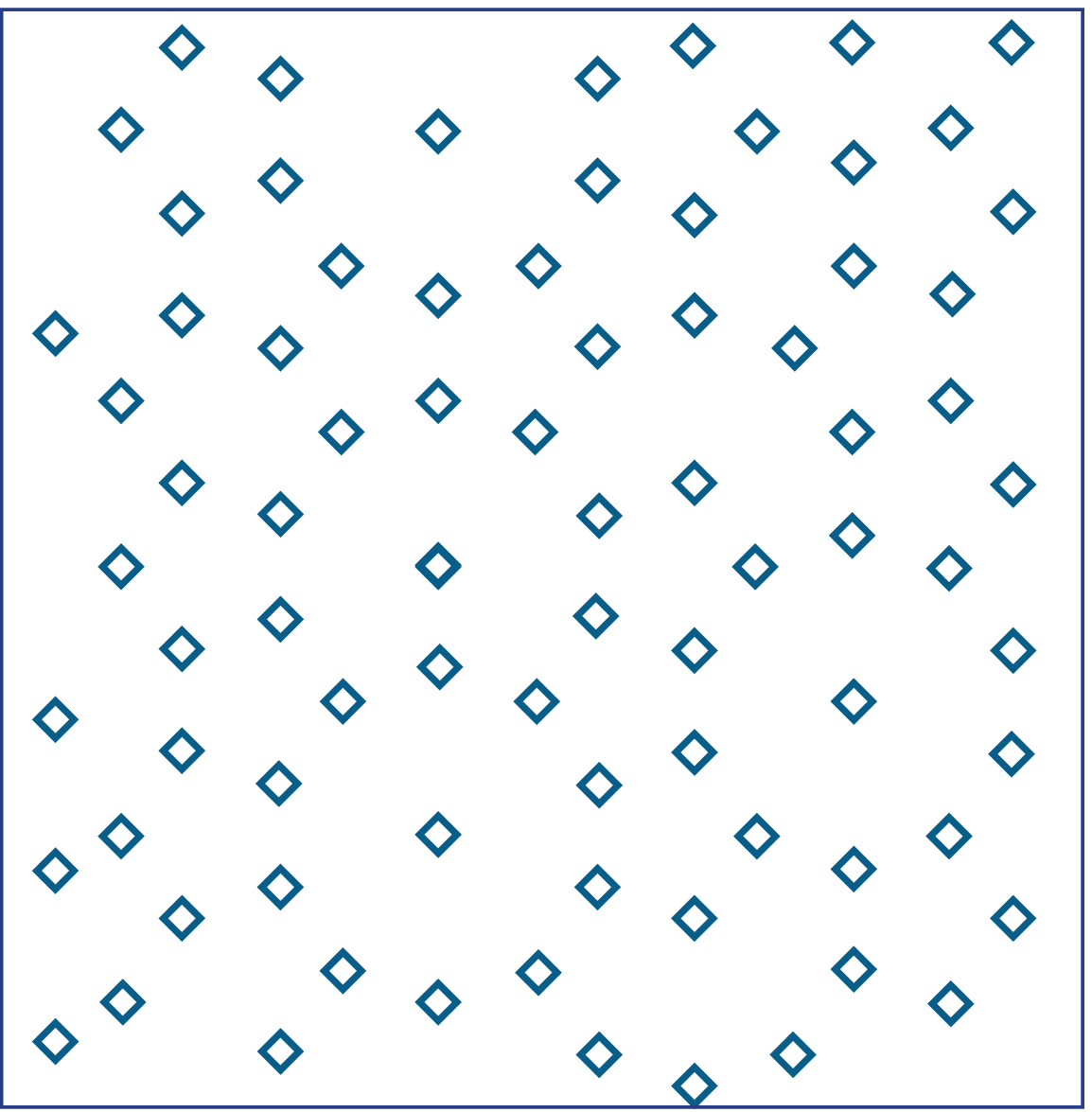}\hfil
\includegraphics[width=0.4\hsize]{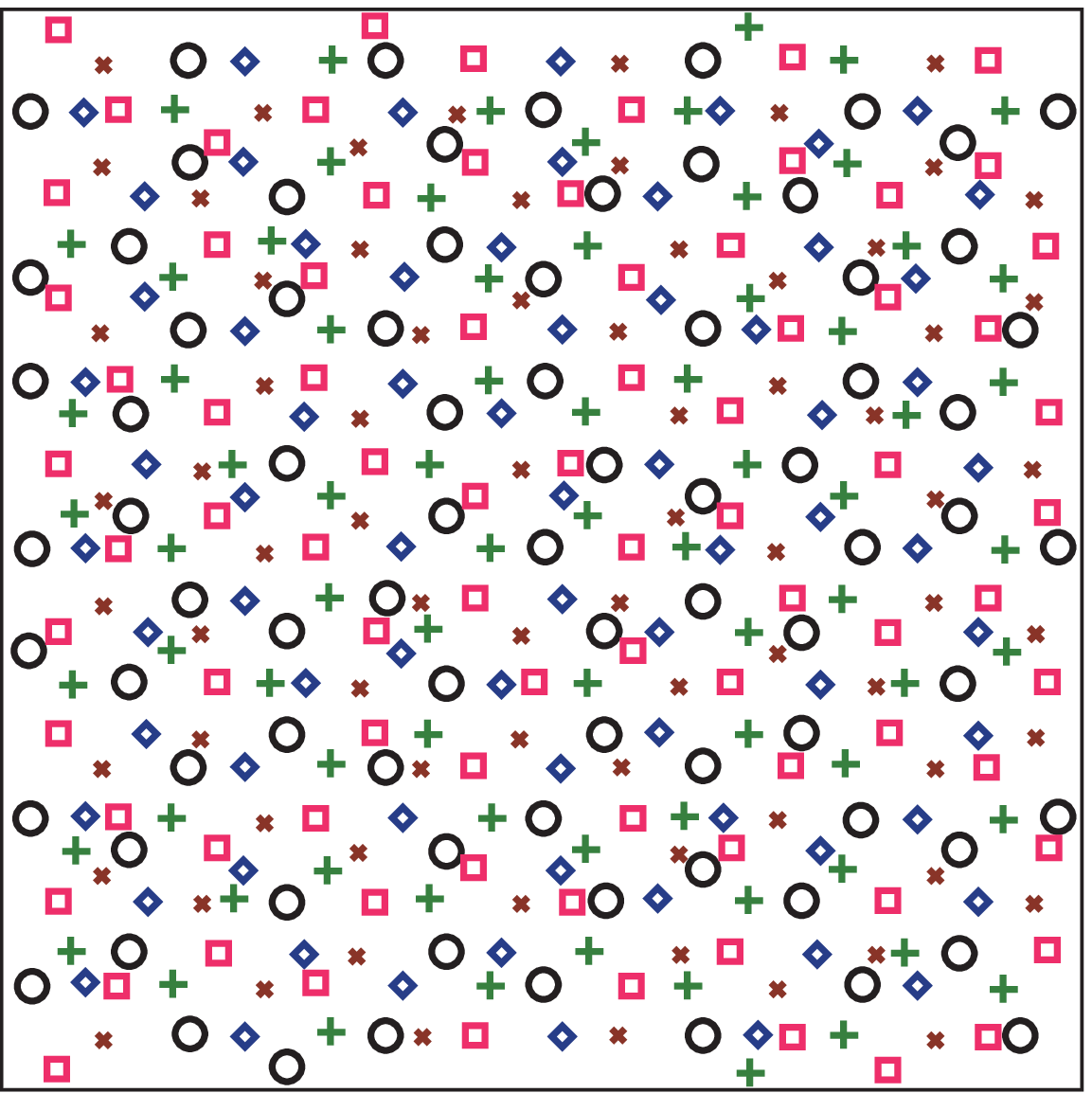}
\end{center}
\vskip6pt
\hbox to\hsize{\hspace*{12pt}\footnotesize
\hfil (a) \hfil\hfil (b)\hfil\hspace*{-2pt}}

\caption{The allowed positions in each layer. (a) Allowed sites at
height $z=3$ are plotted. (b) All layers are plotted: points for $z=1$ are
represented by black boxes; for $z=2$ by red circles; for $z=3$ by
blue diamonds; for $z=4$ by green crosses; for $z=5$ by brown points.}
\end{figure}

Therefore, $\cal L$ is another example of a decagonal quasicrystal.
Decagonal quasicrystals exist in nature, and have been extensively studied
experimentally and theoretically \cite{YaIs,Steurer,SteKuo,
CoWidom,SteH,CHSt}. From Steurer's review \cite{Steurer}, one finds
translational periods along the tenfold $z$-axis varying from 4$\,$\AA\ to
16$\,$\AA\ for different alloys, which may allow 2 layers \cite{CoWidom},
or even 5 or more layers, within a period. 

A situation somewhat similar to ours is found by Ben-Abraham, Lerer and
Snapir \cite{ALS}, who find that the projection of a 6-d lattice to 3d for
a certain choice of bases produces a quasicrystal lattice which is also
periodic in the $z$-direction. Their model has six-fold rather than
five-fold symmetry.

The projection of ${\mbox{\mymsbm Z}}^5\to{\mbox{\mymsbm R}}^3$ is
well-known \cite{YaIs,SteH,Kramer2}. However, interpreting the 3-d
quasicrystal lattice $\cal L$ as overlapping polytopes $\cal P$ provides a
more systematic way to understand this lattice. We find that each unit
cell has 26 sites, sharing the four interior sites with its neighbors. The
lattice is periodic in the $z$-direction with period 5, and quasiperiodic
in the
$xy$-directions. Unlike the
projection of ${\mbox{\mymsbm Z}}^5\to{\mbox{\mymsbm R}}^2$, which
undergoes drastic change in behavior (no inflation and deflation rules) if
$c\ne0$, the projections into 3d are in the same class for $c=0$ and
$c\ne0$.

Since real quasicrystals have icosahedra or triacontahedra as unit cells,
which are projections of a six-dimensional hypercube to a 3-d
space \cite{LRK}, the above method perhaps can also be used to determine
all possible overlappings and their frequencies.
\ack
We are most grateful to Dr.\ M.\ Widom, Dr.\ C.\ Richard, and Dr.\ M.\
Baake for providing us with many useful references. Comments on our
manuscript by Dr.\ M.\ Widom and Dr.\ M.\ Baake are also much
appreciated.


\end{document}